\documentclass[showpacs,oneside,twocolumn,prl,amsmath,amssymb,fleqn,superscriptaddress]{revtex4-1}
\usepackage{cases}
\usepackage{amsmath}
\usepackage{amssymb}
\usepackage{amsfonts}
\usepackage{amssymb}
\usepackage{dcolumn}
\usepackage{bm}
\usepackage{bbm}
\usepackage{graphicx}
\usepackage{xcolor}
\usepackage{array}
\usepackage{subfigure}
\usepackage[none]{hyphenat}

\begin{document}
\title{Observation of Time-domain Rabi Oscillations in the Landau-Zener Regime with a Single Electronic Spin}

\author{Jingwei Zhou}
\affiliation{Hefei National Laboratory for Physics Sciences at
Microscale and Department of Modern Physics, University of Science
and Technology of China, Hefei, 230026, China}
\author{Pu Huang}
\affiliation{Hefei National Laboratory for Physics Sciences at
Microscale and Department of Modern Physics, University of Science
and Technology of China, Hefei, 230026, China}
\author{Qi Zhang}
\affiliation{Hefei National Laboratory for Physics Sciences at
Microscale and Department of Modern Physics, University of Science
and Technology of China, Hefei, 230026, China}
\author{Zixiang Wang}
\affiliation{Hefei National Laboratory for Physics Sciences at
Microscale and Department of Modern Physics, University of Science
and Technology of China, Hefei, 230026, China}
\author{Tian Tan}
\affiliation{Hefei National Laboratory for Physics Sciences at
Microscale and Department of Modern Physics, University of Science
and Technology of China, Hefei, 230026, China}
\author{Xiangkun Xu}
\affiliation{Hefei National Laboratory for Physics Sciences at
Microscale and Department of Modern Physics, University of Science
and Technology of China, Hefei, 230026, China}
\author{Fazhan Shi}
\affiliation{Hefei National Laboratory for Physics Sciences at
Microscale and Department of Modern Physics, University of Science
and Technology of China, Hefei, 230026, China}
\author{Xing Rong}
\affiliation{Hefei National Laboratory for Physics Sciences at
Microscale and Department of Modern Physics, University of Science
and Technology of China, Hefei, 230026, China}

\author{S. Ashhab}
\affiliation{Advanced Science Institute, RIKEN, Wako-shi, Saitama 351-0198, Japan}

\author{Jiangfeng Du}
\altaffiliation{djf@ustc.edu.cn}
\affiliation{Hefei National Laboratory for Physics Sciences at
Microscale and Department of Modern Physics, University of Science
and Technology of China, Hefei, 230026, China}

\begin{abstract}
Under resonant conditions, a long sequence of landau-zener transitions can lead to Rabi oscillations.  Using a nitrogen-vacancy (NV) center spin in diamond, we investigated the interference between more than 100 Landau-Zener processes. We observed the new type of Rabi oscillations of the electron spin resulting from the interference between successive Landau-Zener processes in various regimes, including both slow and fast passages. The combination of the control techniques and the favorable coherent properties of NV centers provides an excellent experimental platform to study a variety of quantum dynamical phenomena.

\end{abstract}


\maketitle

The phenomenon of Rabi oscillations, first studied in 1937
\cite{Rabi}, occurs in almost any quantum system under the influence
of resonant external driving and is at the heart of various
spectroscopic techniques. Landau-Zener (LZ) transitions, first
studied in 1932 \cite{Landau,Zener,Stuckelberg,Majorana}, are
intriguing phenomena that are ubiquitous in quantum systems,
typically occurring when two energy levels of a quantum system
undergo an avoided crossing.

Under suitable conditions LZ transitions can be treated as quantum
coherent processes, and multiple such processes can interfere
constructively or destructively \cite{Stuckelberg}. Depending on
the details of the interference, a long and regular sequence of LZ
processes can exhibit resonance behavior similar to that seen in
Rabi oscillations under weak driving conditions. The oscillations obtained in
the case of constructive interference can therefore be seen as a
manifestation of Rabi oscillations in the regime of ultrastrong
driving \cite{Shevchenko}. Moreover, the interference between LZ
processes gives rise to a number of novel features distinct from
the case of weak driving. Particularly interesting is the fact
that the patterns of the resonance lines vary drastically
depending on whether each passage through the avoided crossing is
slow or fast \cite{Shevchenko}.

The experimental observation of Rabi oscillations in the LZ regime
requires stringent conditions. The coherence time of the quantum
system is required to be long enough to allow the coherent interference
between multiple LZ processes, and the control fields are required to be accurate
and stable in order to precisely adjust the quantum phases that
govern the interference effects. In addition, the time resolution
of the measurement needs to be high enough to allow the monitoring of the dynamics on short
timescales.

With the recent advances in various quantum systems
\cite{Ladd,Buluta}, a number of experiments were able to
demonstrate the controlled interference of LZ transitions.
Interference between two LZ transitions has been observed in
gaseous molecules \cite{Mark}, semiconductor-based quantum dots
\cite{Petta}, NV centers \cite{PuHuang}, and atoms in optical lattices \cite{Kling}. Evidence
for various interference effects involving multiple LZ transitions
has been observed in the steady-state behavior of continuously driven
superconducting qubits \cite{Oliver,Sillanpaa} and NV centers \cite{Childress}.
However, although those steady-state behaviors have been reported, the experimental observation
of the abundant time-domain evolutions remains elusive due to more stringent conditions required.

Here by precise quantum control of the NV center in diamond at room temperature, we experimentally observed
time-domain Rabi oscillations resulting from the interference between more than 100 LZ
processes. Our setup allows us to monitor the dynamics on both
short and long timescales. On short timescales we clearly see the
step-like dynamics that is characteristic of LZ transitions, while
on long timescales we see oscillations that are characteristic of
Rabi oscillations in a resonantly driven system. We further vary
the parameters of our control fields in order to vary the degree
of interference between successive LZ transitions. Our results
therefore constitute a direct demonstration of fully controllable
multi-transition interference in its most general form. All of our
experimental results agree well with numerical simulations using
parameters that are obtained from standard characterization
techniques.
\begin{figure}[htbp]
\centering
\includegraphics[width=1\columnwidth]{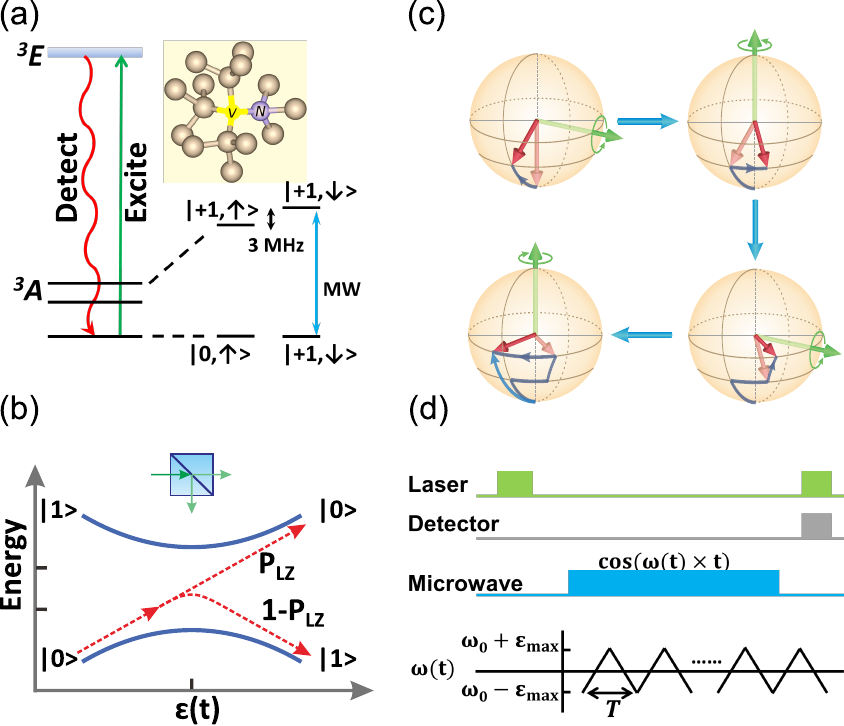}
\caption{(color online). (a) Schematic diagram of an NV center in the
diamond lattice and the energy level diagram of the NV center.
(b) A single Landau-Zener transition: when the energy levels undergo
an avoided crossing, the quantum state undergoes a mixing process
between the ground and excited states. This process is analogous
to the operation of a beam splitter in optics experiments. (c)A single driving period contains
four elementary rotations, which can be combined to produce a
single effective rotation. On short timescales, one can see that
each period involves four different rotations. On long timescales,
the dynamics is described by the repeated application of the
single-period rotation, as is the case for Rabi oscillations. (d) The pulse sequence used to drive Rabi oscillations in the LZ
regime.}
\label{fig1}
\end{figure}

A NV center consists of a substitutional N atom with an adjacent
vacant site (V) in the diamond crystal lattice. It has an electron spin-1 ground
state with three sublevels, $|m_s=0\rangle$ and
$\left|m_s=\pm1\right\rangle$, quantized along the [111] crystal
axis [Fig.\  1(a)]. Quantum logic gates
\cite{Jelezko1,Jelezko2}, quantum entanglement generation
\cite{Neumann}, single-shot readout \cite{Neumann2,Buckley,Robledo} and quantum algorithms \cite{Fazhan,Hanson} have been
realized in such systems. These achievements make NV centers in
diamond a promising candidate as a platform for building quantum
computers. Furthermore, with their favorable coherence properties
and atomic-scale size, NV centers have also been used for sensing magnetic field with nanometer spatial resolution
\cite{Balasubramanian1,Maze}.
\begin{figure}[htbp]
\begin{center}
\includegraphics[width=1\columnwidth]{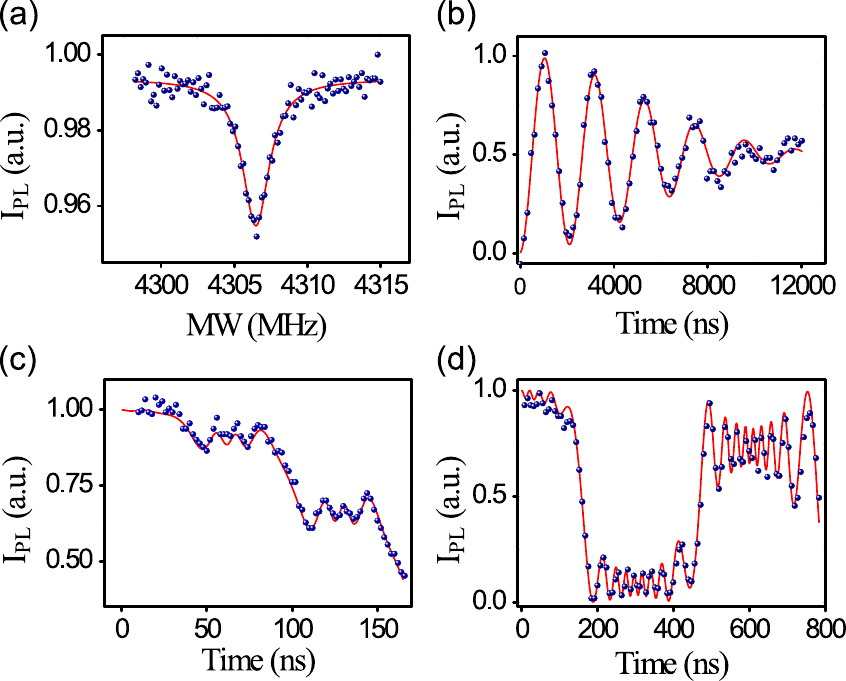}
\end{center}
\caption{(color online). (a) The optically detected magnetic resonance (ODMR)
spectrum for the transition between the states $|0\rangle$ and
$|1\rangle$ of the NV center (blue dots). The red curve is a fit
with a single Lorentzian peak. (b) Results of a standard Ramsey experiment of the NV
center. The results are fitted to the function $\exp[-(t/T_{2}^{\ast})^{2}]\cos(2\pi\delta_f t)$ ($\delta_f = 0.56~$MHz), where the oscillations are due to the microwave detuning. (c,d) The occupation probability of the state $|0\rangle$ for a
double-passage experiment in the fast- and slow-passage regimes.
}
\label{fig2}
\end{figure}

The sample we used is an isotopically pure
${}^{12}$C ($>$99.9\%) diamond sample grown by chemical vapor deposition. When an externally applied magnetic field is aligned with the principal axis of the NV center, the Hamiltonian for the NV center is:
\begin{equation}
\label{eqNVHamiltonian} H = D S_{z}^2 + \gamma_e B S_{z} +
A_{zz}I_{z}S_{z}.
\end{equation}
Here $D = 2.87~$GHz, $\gamma_e = 2.8~$MHz/G, $A_{zz} = -3.05~$MHz are parameters for the zero-field splitting, the Zeeman interaction, and the hyperfine interaction of the electron spin with the ${}^{15}$N nuclear spin ($I = 1/2$) of the NV center, respectively. In order to realize a simple two-level system, we apply a magnetic field of $\thicksim$ 510 G. This field lifts the
degeneracy between the states $\left|m_s=\pm1\right\rangle$. Furthermore, the
nitrogen nuclear spins are polarized by optical pumping
\cite{Jacques}, such that $I_z=-1/2$ throughout the experiment. We use the states $|m_s=0\rangle$ and $|m_s=+1\rangle$ for the qubit's states $|0\rangle$ and $|1\rangle$, respectively. 


To induce transitions between the states $|0\rangle$ and
$|1\rangle$, microwave (MW) fields are generated using a local
oscillator mixed with an arbitrary-waveform generator and irradiated to the NV center
via a coplanar waveguide. The form of the variable-frequency MW signal is $2\Delta
\cos[\omega(t)\cdot t]$. Since $\omega(t)$ remains close to $\omega_0$, we define the
difference $\varepsilon(t)=\omega(t)-\omega_0$, i.e. $\varepsilon(t)$ is the offset between the microwave
signal frequency and the resonance frequency of the $|0\rangle
\leftrightarrow |1\rangle$ transition. If $\omega(t)$ is
varied slowly compared to $\omega_{0}$ and $\Delta\ll\omega_0$, we
can transform the Hamiltonian to a frame rotating with frequency
$\omega_0$ and apply a rotating-wave approximation to simplify the Hamiltonian:
\begin{equation}
\label{eqQubitHamiltonian} H =
\frac{\varepsilon(t)}{2}\sigma_{z}+\frac{\Delta}{2}\sigma_{x}.
\end{equation}
Here we use a triangle wave for the function
$\varepsilon(t)$:
{
\begin{displaymath}
\varepsilon(t) = \left\{ \begin{array}{ll}
-\varepsilon_{\rm m}+\frac{4\varepsilon_{\rm m}(t-nT)}{T}, & \textrm{$t \in(n,n+\frac{1}{2})T$}\\
\varepsilon_{\rm m}-\frac{4\varepsilon_{\rm m}(t-[n+1/2]T)}{T}, & \textrm{$t \in(n+\frac{1}{2},n+1)T$}\\
\end{array} \right.
\end{displaymath}
}
where $2\varepsilon_{\rm m}$ and $T$ are, respectively, the
range and period of the function $\varepsilon(t)$, and $n=0,~1,~2,\cdots$. When the system is initialized in the state  $|0\rangle$ and
$\varepsilon(t)$ is swept through the value zero, a simple LZ
transition occurs. The probability of the transition from
$|0\rangle$ to $|1\rangle$ is given by the LZ formula [Fig.\ 1(b)]
\begin{equation}
\label{eqLZProbability} P_{|0\rangle \rightarrow
|1\rangle}=1-P_{\rm LZ} =
1-\exp(-\frac{\pi}{2}\frac{\Delta^{2}}{v})
\end{equation}
where $v$ is the sweep rate of the diabatic energy difference at
the crossing. For the triangle wave used in our experiment the sweep rate is $4\varepsilon_{\rm
m}/T$.

Firstly, we characterize the parameters of the system. The optically detected
magnetic resonance spectrum for the $|0\rangle \rightarrow
|1\rangle$ transition is plotted in Fig.\ 2(a). The appearance of a
single ESR resonance peak confirms the polarization of the nearby
nuclear spin.  The free-induction-decay signal, shown in
Fig.\ 2(b), gives a rather long dephasing time of
$T_{2}^{\ast}=6.56 \pm 0.17~\mu$s and spin-echo measurements (data not shown) give approximately $T_{2}=140~ \mu$s.
For a slow sweep, the system
transforms adiabatically from $|0\rangle$ to $|1\rangle$. As we
make the sweep time shorter, $P_{\rm LZ}$ follows the
exponential-decay function given in Eq.~(\ref{eqLZProbability}).
Using this experiment we can extract the value of $\Delta$, and the
values obtained this way always agree with the values obtained
using resonant driving (see Supplemental Material). In Fig.\ 2(c,d), we show the results for
double-passage LZS interferometry in the fast- and slow-passage
regimes corresponding to $P_{\rm LZ}$ close to 1 and 0,
respectively. This short-time dynamics clearly shows the step-like
LZ transitions and is in good agreement with the results of
numerical simulations based on experimentally obtained parameters.

We now consider the effects of strong, periodic
driving of the NV center.
Landau-Zener-St\"uckelberg (LZS)
interference can be understood using an analogy with Mach-Zehnder
interferometry in an optical setup \cite{Oliver}.
This analogy provides a good understanding of the LZS setup
when dealing with one or two LZ processes.
For analyzing the
resonance phenomena associated with continuous driving of the
system, a geometric picture is more intuitive
\cite{Sillanpaa,Ashhab}.
When the two-level quantum system is driven through the avoided crossing
region, the ground and excited states are mixed together in a
process that can be described by the following matrix (See also Supplementary Material)

\begin{equation}
\label{eqLZMixingMatrix} N = \left( \begin{array}{cc} \alpha & -\gamma^* \\
\gamma & \alpha^*
\end{array}
\right)
\end{equation}
where $\alpha=\sqrt{1-P_{\rm LZ}} e^{-i\varphi}$ and
$\gamma=\sqrt{P_{\rm LZ}}$, with $\varphi$ the geometric phase
factor \cite{Shevchenko}.

In the geometric picture, the LZ mixing processes is represented
by a rotation about an axis that depends on the various parameters
described above. Between LZ processes the system is moved away
from the avoided crossing and brought back some time later, such
that a phase accumulates between the two quantum states. This
phase is given by the integral over time of the energy separation
between the two energy levels $2\zeta=\int(E_e-E_g)dt$, and the
corresponding quantum process $U$ is represented by a rotation
about the $z$ axis.
A full driving cycle is therefore composed of four rotations alternating
between two different axes [Fig.\ 1(c)]. This sequence of four rotations is
equivalent to a single rotation ($G_1=U_2NU_1N$); depending on the
various angles involved, the axis for the net rotation can point
in any direction. Specifically, the matrix $G_1$ describing the
net rotation has the same form as Eq.~(\ref{eqLZMixingMatrix}),
but with parameters $\alpha$ and $\gamma$ that can be lengthy in
general \cite{Shevchenko}.

By stringing a large number of these rotations together, we
obtain a resonance effect. If the axis of the single-period net
rotation points in the $xy$ plane, the resonance condition is
satisfied. Even if the mixing between the ground and the excited
state in a single driving period is small, the rotations of
multiple periods will add up constructively, and we obtain full
oscillations between the states $|0\rangle$ and $|1\rangle$. If
the axis of the single-period net rotation points outside the $xy$
plane, there can be partial, but not full, conversion between the
two states. Two opposite limits are of particular interest: fast
(almost sudden, $1-P_{\rm LZ}\ll1$) and slow (almost
adiabatic, $P_{\rm LZ}\ll1$) passages. In both cases, we
observe dynamics that reflects LZ processes associated with the
passage through the avoided crossing as well as
Rabi-oscillation dynamics on large timescales(Fig.\ 3).
\begin{figure}[htbp]
\begin{center}
\includegraphics[width=1\columnwidth]{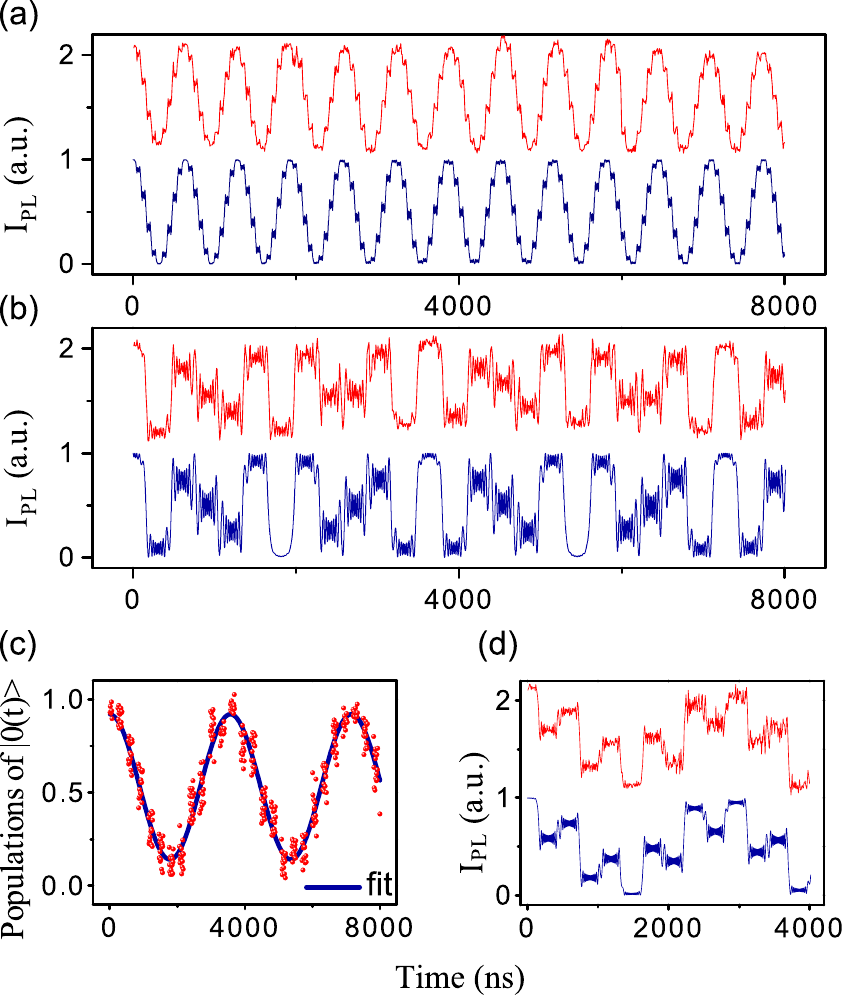}
\end{center}
\caption{(color online). Rabi oscillations resulting from the interference of many
LZ processes.
(a) \textit{fast-passage regime}. The measured (pink line) and numerically simulated (blue line) occupation probabilities $P_{0}$ of the state as a function of time. The experimental data here and those in (b) and (d) are offset in y axis for visibility. The Rabi Frequency is about $1.49~$MHz, which is consistent with numerical simulations.
(b,~c) \textit{slow-passage regime}. Oscillations plotted the diabatic basis and in the adiabatic basis.
(d) Dynamics in the regime of an intermediate sweep rate with partially destructive
interference.}
  \label{fig3}
\end{figure}
\begin{figure}[htbp]
\begin{center}
\includegraphics[width=1\columnwidth]{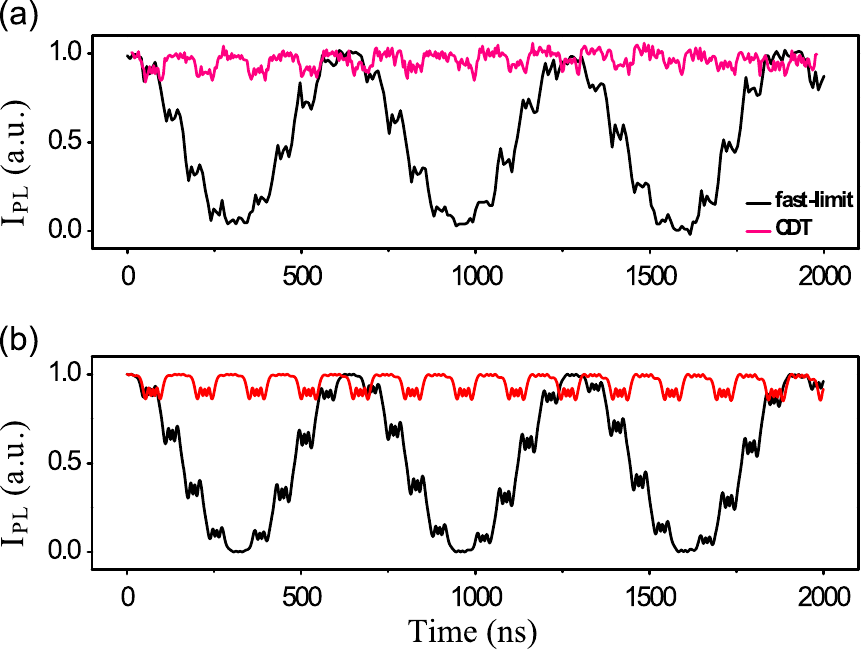}
\end{center}
\caption{(color online). Constructive versus destructive interferences in a single
driving period. The parameters are the same as in Fig.\ 3(a),
except for the sweep period being $T=149~$ns in the case
of destructive interference (pink line). The experimental results
are shown in (a), while the results of numerical simulations are
shown in (b). The case of destructive interference corresponds to
the phenomenon of coherent destruction of tunneling.}
\label{fig4}
\end{figure}

For the data plotted in Fig.\ 3(a), we set $\Delta= 5.57$
MHz, $\varepsilon_{\rm m}= 100$ MHz and $T$=128 ns,
which gives $\varepsilon_{\rm m}/\Delta\approx 18$ (well in the
LZ regime) and $P_{\rm LZ}\approx 0.91$ (fast-passage regime). We
observe coherent oscillations that result from the quantum
interference of a sequence containing over 100 LZ processes. These
oscillations are observed up to about 8 $\mu$s without any
appreciable steady decay. For the data plotted in Fig.\ 3(b), we set
$\Delta= 9.60$ MHz, $\varepsilon_{\rm m}= 50.4$ MHz and $T$= 606 ns, which gives $\varepsilon_{\rm
m}/\Delta\approx 5$ (LZ regime) and $P_{\rm LZ}\approx 0.065$
(slow-passage regime). When we plot the data in the basis of the states
$|0\rangle$ and $|1\rangle$, we can clearly see the steps
associated with LZ processes. However, the long-time dynamics looks rather irregular and the oscillations are not easily visible. In order to show the oscillations more
clearly, we transform the results to the adiabatic basis, as shown
in Fig.\ 3(c). There we only keep data points for
which $|\varepsilon(t)|>3\Delta$, such that the difference between
the diabatic and adiabatic bases is under 3\% and the
transformation of the state populations between the two bases can
be done straightforwardly with negligible errors. The
Rabi-oscillation pattern is now clearly recognizable. The
observation of Rabi oscillations in this case is possibly more
impressive than that in the fast-passage regime. Every passage
through the avoided crossing is now almost adiabatic and
drastically transforms the character of the quantum state.
Nevertheless, quantum coherence is preserved and interference of
over 15 LZ processes is observed with no discernible decay. In Fig.\ 3(d) we
show the data that correspond to the regime of intermediate sweep
speed ($\Delta= 5.84$ MHz, $\varepsilon_{\rm
m}= 100$ MHz and $T$=592 ns, which gives $P_{\rm
LZ}=0.61$) and an intermediate degree of constructive interference
between successive LZ processes. Specifically, the interference
between two LZ processes within a single driving period is
somewhat destructive, as can be seen from the fact that each step
goes in the opposite direction from the previous one.
The effect of varying the degree of constructive interference is
also illustrated in Fig.\ 4. Taking the parameters of Fig.\ 3(a) and
changing $T$ from 128 ns to 149 ns, the constructive interference
between successive LZ processes is replaced by completely
destructive interference. This situation corresponds to the
so-called coherent destruction of tunneling \cite{Grossmann},
where the $|0\rangle \rightarrow |1\rangle$ transition is
suppressed not because the driving is non-resonant, but because
the parameters are chosen such that the single-period net
rotation gives the identity operation (i.e.~it has a zero rotation
angle).

In summary, using a single electron spin in diamond, we have
observed Rabi oscillations resulting from the quantum interference
of more than 100 LZ processes in various regimes ranging from slow to
fast passage and from constructive to destructive interference
within a single driving period. The strong driving of the NV
center has also allowed us to obtain the values of system
parameters using alternative techniques, and the values extracted
using multiple techniques have consistently agreed with one
another. The experimental observation of the quantum oscillations in the LZ regime verifies
the rich dynamics of a fundamental paradigm. The combination of precise control techniques with the favorable coherent properties of NV centers provide an excellent experimental platform with great potential in various fields, ranging from quantum information processing to the simulation of various dynamical phenomena in chemical and biology physics.

We thank J. Wrachtrup from University of Stuttgart, Germany for providing the isotopically purified diamond sample, Yiqun Wang in Suzhou Institute of Nano-Tech and Nano-Bionics for fabricating the coplanar waveguide, and Changkui Duan for helpful discussions. This work was supported by the National Key Basic Research Program of China (Grant No. 2013CB921800), the National Natural Science Foundation of China (Grant Nos. 11227901, 11275183, 91021005 and 10834005), the `Strategic Priority Research Program (B)' of the CAS (Grant No. XDB01030400) and the Fundamental Research Funds for the Central Universities.


\begin{references}
\bibitem{Rabi} I. I. Rabi, Phys. Rev. \textbf{51}, 652 (1937).
\bibitem{Landau} L. Landau, Phys. Z. Sowjetunion \textbf{1}, 88 (1932); \textit{ibid.} \textbf{2}, 46 (1932).
\bibitem{Zener} C. Zener, Proc. R. Soc. Lond. A \textbf{137}, 696 (1932).
\bibitem{Stuckelberg} E. C. G. Stueckelberg, Helv. Phys. Acta \textbf{5}, 369 (1932).
\bibitem{Majorana} E. Majorana, Nuovo Cimento \textbf{9}, 43 (1932).
\bibitem{Shevchenko} S. N. Shevchenko, S. Ashhab, and F. Nori, Phys. Rep. \textbf{492}, 1 (2010).
\bibitem{Ladd} T. D. Ladd, F. Jelezko, R. Laflamme, Y. Nakamura, C. Monroe, and J. L. O'Brien, Nature \textbf{464}, 45 (2010).
\bibitem{Buluta}I. Buluta, S. Ashhab, and F. Nori, Rep. Prog. Phys. \textbf{74}, 104401 (2011).
\bibitem{Mark} M. Mark, T. Kraemer, P. Waldburger, J. Herbig, C. Chin, H.-C. N\"agerl, and R. Grimm, Phys. Rev. Lett. \textbf{99}, 113201 (2007).
\bibitem{Petta} J. R. Petta, H. Lu, and A. C. Gossard, Science {\bf 327}, 669 (2010).
\bibitem{PuHuang} P. Huang, J. Zhou, F. Fang, X. Kong, X. Xu, C. Ju, and J. Du, Phys. Rev. X \textbf{1} 011003 (2011).
\bibitem{Kling} S. Kling, T. Salger, C. Grossert, and M. Weitz, Phys. Rev. Lett. \textbf{105}, 215301 (2010).
\bibitem{Oliver} W. D. Oliver, Y. Yu, J. C. Lee, K. K. Berggren, L. S. Levitov, and T. P. Orlando, Science \textbf{310}, 1653 (2005).
\bibitem{Sillanpaa} M. Sillanp\"a\"a, T. Lehtinen, A. Paila, Yu. Makhlin, and P. Hakonen, Phys. Rev. Lett. \textbf{96}, 187002 (2006).
\bibitem{Childress} L. Childress and J. McIntyre , Phys. Rev. A \textbf{82}, 033839 (2010).
\bibitem{Fuchs3} G. D. Fuchs, G. Burkard, P. V. Klimov, and D. D. Awschalom, Nature Phys. \textbf{7}, 789 (2009).
\bibitem{Fuchs} G. D. Fuchs, V. V. Dobrovitski, D. M. Toyli, F. J. Heremans, and D. D. Awschalom, Science \textbf{326}, 1520 (2009).
\bibitem{Jelezko1} F. Jelezko, T. Gaebel, I. Popa, A. Gruber, and J. Wrachtrup, Phys. Rev. Lett. \textbf{92}, 076401 (2004).
\bibitem{Jelezko2} F. Jelezko, T. Gaebel, I. Popa, M. Domhan, A. Gruber, and J. Wrachtrup, Phys. Rev. Lett. \textbf{93}, 130501 (2004).
\bibitem{Neumann} P. Neumann, N. Mizuochi, F. Rempp, P. Hemmer, H. Watanabe, S. Yamasaki, V. Jacques, T. Gaebel, F. Jelezko, J. Wrachtrup, Science \textbf{320}, 1326 (2008).
\bibitem{Buckley} B. B. Buckley, G. D. Fuchs, L. C. Bassett, D. D. Awschalom, Science \textbf{330}, 1212 (2010).
\bibitem{Neumann2} P. Neumann, J. Beck, M. Steiner, F. Rempp, H. Fedder, P. R. Hemmer, J. Wrachtrup and F. Jelezko, Science \textbf{329}, 542 (2010).
\bibitem{Robledo} L. Robledo, L. Childress, H. Bernien, B. Hensen, Paul F. A. Alkemade, R. Hanson, Nature \textbf{477}, 574 (2011).
\bibitem{Fazhan} F. Shi, X. Rong, N. Xu, Y. Wang, J. Wu, B. Chong, X. Peng, J. Kniepert, R. Schoenfeld, W. Harneit, M. Feng, and J. Du, Phys. Rev. Lett. \textbf{105}, 040504 (2010).
\bibitem{Hanson} T. van der Sar, Z. H. Wang, M. S. Blok, H. Bernien, T. H. Taminiau, D. M. Toyli, D. A. Lidar, D. D. Awschalom, R. Hanson and V. V. Dobrovitski, Nature \textbf{484}, 82 (2012).
\bibitem{Balasubramanian1} G. Balasubramanian, I. Y. Chan, R. Kolesov, M. AlHmoud, J. Tisler, C. Shin, C. Kim, A. Wojcik, Philip R. Hemmer, A. Krueger, T. Hanke, A. Leitenstorfer, R. Bratschitsch, F. Jelezko and J. Wrachtrup, Nature \textbf{455}, 648 (2011).
\bibitem{Maze} J. R. Maze, P. L. Stanwix, J. S. Hodges, S. Hong, J. M. Taylor, P. Cappellaro, L. Jiang, M. V. Gurudev Dutt, E. Togan, A. S. Zibrov, A. Yacoby, R. L. Walsworth and M. D. Lukin, Nature \textbf{455}, 644 (2011).
\bibitem{Jacques} V. Jacques, P. Neumann, J. Beck, M. Markham, D. Twitchen, J. Meijer, F. Kaiser, G. Balasubramanian, F. Jelezko, and J. Wrachtrup, Phys. Rev. Lett.  \textbf{102}, 057403 (2009).
\bibitem{Ashhab} S. Ashhab, J. R. Johansson, A. M. Zagoskin, and F. Nori, Phys. Rev. A \textbf{75}, 063414 (2007).
\bibitem{Grossmann} F. Grossmann, T. Dittrich, P. Jung, and P. H\"anggi, Phys. Rev. Lett. \textbf{67}, 516 (1991).

\clearpage

\end{references}
\end{document}